\documentclass[12pt,a4paper]{elsarticle}

\usepackage{amssymb}
\usepackage{xcolor}
\usepackage[para]{footmisc}
\usepackage{comment}
\usepackage{datetime}

\usepackage{hyperref}
\usepackage{xcolor}
\usepackage{scalerel}
\usepackage{tikz}
\usepackage{lipsum}

\makeatletter
\def\ps@pprintTitle{%
 \let\@oddhead\@empty
 \let\@evenhead\@empty
 \def\@oddfoot{\centerline{\thepage}}%
 \let\@evenfoot\@oddfoot}
\makeatother

\begin{document}

\begin{frontmatter}

\title{Autonomous Shuttle-as-a-Service (ASaaS): \\Challenges, Opportunities, and Social Implications}



\author{Antonio Bucchiarone\href{https://orcid.org/0000-0003-1154-1382}{\includegraphics[scale=.06]{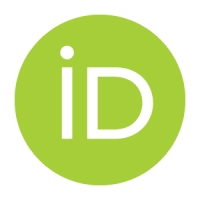}}, Sandro Battisti \href{https://orcid.org/0000-0002-7818-1230}{\includegraphics[scale=.06]{orcid.png}}, Annapaola Marconi\href{https://orcid.org/0000-0001-8699-7777}{\includegraphics[scale=.06]{orcid.png}}}
\address{Fondazione Bruno Kessler, Trento, Italy}
\address{\{bucchiarone,s.battisti,marconi\}@fbk.eu}
\author{Roberto Maldacea}
\address{I-Mobility Garage, Verona, Italy}
\address{roberto.maldacea@nevmobility.eu}

\author{Diego Cardona Ponce }
\address{Navya Tech, Lion, France}
\address{diego.cardona-ponce@navya.tech}
\address{}{}{}
\address{}{}{}
\address{\textbf{\today}}



\begin{abstract}

Modern cities are composed of complex socio-technical systems that exist to provide services effectively to their residents and visitors. In this context, \textit{smart mobility} systems aim to support the efficient exploitation of the city transport facilities as well as sustainable mobility within the urban environment. \textit{People} need to travel quickly and conveniently between locations at different scales, ranging from a trip of a few blocks within a city to a journey across cities or further. At the same time, \textit{goods} need to be timely delivered considering the needs of both the users and the businesses. While most of the mobility and delivery solutions can cover significant distances and multiple requests, they suffer when the requests come from the growing neighborhoods and hard-to-reach areas such as city centers, corporate headquarters, and hospitals. In the last few years, several cities indicated interest in using Autonomous Vehicles (AV) for the \textit{“last-mile"} mobility services. With them, it seems to be easier to get people and goods around using fewer vehicles.  In this context, \textit{Autonomous Shuttles} (AS) are beginning to be thought of as a new mobility/delivery service into the city center where narrow streets are not easily served by traditional buses. They allow them to serve critical areas with minimal new infrastructure and reducing noise and pollution. The goal of this article is to present an innovative vision on the introduction of the Autonomous Shuttles-as-a service (ASaaS) concept as the key pillar for the realization of \textit{innovative and sustainable proximity mobility}. Through a set of real application scenarios, we present our view, and we discuss a set of challenges, opportunities, and social implications that this way to reimage the mobility of the future introduces.

\end{abstract}


\end{frontmatter}

\noindent
\section{Introduction}
The mobility of people and goods is in the center of transportation planning and decision-making of the cities of the future\footnote{\url{https:/urban.jrc.ec.europa.eu/thefutureofcities/mobility/\#the-chapter}}. Smart cities are prioritizing the walking, cycling, and public transport based on electric vehicles and other efficient shared mobility, as well as their interconnectivity, to accelerate the transition to zero-emission vehicles and maximize climate and air quality benefits. To make this possible, cities are introducing mechanisms to discourage the use of cars, single-passenger taxis, and other oversized vehicles transporting one person while are introducing mechanisms to encourage citizens to change their behaviors and to make sustainable mobility habits daily.

Recently, cities are investing in the infrastructure and technology necessary to support a connected, multi-modal transit network that includes shared electric \textit{Autonomous Vehicles} (AVs) intending to replace less efficient bus lines \cite{33}. AVs are capable of moving without the full control of humans and focusing on sensing their environment and automate some aspects of safety, such as steering or braking, without human input.

There are many advantages for citizens and involved stakeholders on the use of autonomous mobility that could be brought by increasing the spread of AV to the worldwide market \cite{37,38}. It can be possible to reduce carbon production and have exploitation of the time spent inside the car since drivers can be able to employ their time to do other activities instead of only driving. Moreover, it is possible to reduce traffic, congestion, and accidents that are mainly caused by driver errors, fatigue, alcohol, or drugs \cite{3,4,5}. However, even though the levels of security provided by such system are expected to improve further, and many tests on urban roads have already been made \cite{7,8}, high level of market penetration of completely autonomous vehicles cannot be guaranteed in the next decade \cite{1,3}. Nevertheless, many research fields study the impact of AV introduction and their interaction with conventional vehicles.

In 2018, the U.S. Department of Transportation had released a policy regarding automated vehicles and their safe integration in the transportation system \cite{1} in which categorization of the levels of autonomy can be found. Some of these levels of automation have already been integrated into cars that are on the market, such as self-parking and crash avoidance features, and many manufacturers are venturing in this direction \cite{2}\footnote{\url{https://www.cbinsights.com/research/autonomous-driverless-vehiclescorporations-list/}} . 

The impact of AV on safety is studied in \cite{7} following models \cite{8} to simulate the human-driven vehicles and the relationship with AV. In the same context, the Surrogate Safety Assessment Model (SSAM) \cite{9} has been introduced to assess potential conflicts of a road network and therefore extract the number of potential conflicts during the simulation. 

The investigation is conducted for signaled intersections and roundabouts. The results proved that with high penetration rates, the AVs improve safety significantly. Other studies have focused directly on the modeling of systems to automating intersection crossing. Automating the crossing at intersections could help in improving the traffic flow and avoid collisions. An example of automated intersection crossing, modeled with multi-agent systems, can be found in \cite{10}. Here, intersections are regulated by a Multi-Agent Autonomous Management (MA-AIM) system, which also exploits blockchain technology.

The impacts of self-driving cars portend significant changes to the transportation ecosystem. Some forecast the end of parking spaces \cite{11}. Others believe that AV will paradoxically increase traffic. Others predict that there will be new classes of traffic problems that occur at scale due to the homogeneity of these transportation systems  \cite{11}. Multiple studies report also doubts about the ability to AVs to improve congestion \cite{13} especially when they are used in the proximity of crowded central urban areas.

The \textit{‘last-mile’} mobility term has been introduced to describe the movement of people and goods from a transportation hub to a final destination (home, workplace, shopping center, institution, hospital, etc.). In the \textit{mobility domain}, the last-mile piece of the total transportation path is considered the least efficient part of an overall journey. This fact occurs for different reasons like public transport does not take us fully to the doorstep; the parking spaces are difficult to find and have a car or bicycle always possible. In the context where city populations keep growing, various \textit{micro-mobility} services\footnote{\url{https://www.sharedmobility.news/five-promises-of-micromobility/}}  (i.e., electric scooters, electric skateboards, shared bicycles, etc.) are emerging intending to reduce the efficiency of the last-mile mobility. However, these micro-services are not well integrated with other mobility solutions provided by a city and usually disappear quickly from the market.

In the \textit{logistics domain}, one of the biggest challenges is the home delivery efficiency. Usually, it is not performant due to the spatial dispersion of the parcels' recipients and the frequency of failed deliveries. These problems led to the introduction of new intelligent transport vehicles with the aim to decrease shipping costs and impact (i.e., drones)  \cite{34,35}.

From this perspective and based on the growth of autonomous shuttles in urban public environments could enable new services to deal with the new challenges posed by large cities, which requires the combination of the mobility of people and goods  \cite{14}. In particular, several pilot experimentations prove significant technology development results, as well as the citizens' acceptance in many cities all over the world, in countries such as Germany, France, Switzerland, Finland, Sweden, The Netherlands, and Estonia, as presented by recent research  \cite{14,15}.

We introduce the concept of Autonomous Shuttles-as-a-service (ASaaS) as a part of the Mobility Mobility-as-a-Service (MaaS) paradigm. MaaS solutions (e.g., MaaS Global: http://maas.global) aim at arranging the most suitable transport solution for their customers' thanks to the costs of an effectively integrated offer of different multi-modal means of transportation. MaaS also foresees radical changes in the business landscape, with a new generation of mobility operators emerging as key actors to manage the increased flexibility and dynamism offered by this new concept of mobility. We claim that, in the first and last-mile mobility of people and goods, the ASaaS could be the right concept because it has the potential of offering various mobility services. These services: (i) are tailored to the traveler needs and preferences, (ii) allow to serve critical areas with minimal new infrastructure and reducing noise and pollution, and (iii) complement the already available  (public and private) mobility services in a city solving the well-known issues of the proximity mobility.

The goal of this article is to present our vision on the introduction of the Autonomous Shuttles-as-a service concept, in particular, as the key pillar for the realization of \textit{innovative and sustainable proximity mobility}. Through a set of real application scenarios, we present our view, and we discuss the set of challenges, opportunities, and social implications that this way to reimage the mobility of the future introduces.

\section{Application Scenarios}
\label{sec:scenario}
There are many application scenarios that can make use of Autonomous Shuttles to perform specific services in a city. In this section, we try to present such applications to obtain a general point of view of the different provided services and the stakeholders involved. At the same time, they will help us to introduce our vision in Section \ref{sec:vision} and to deepen their social implications.

\subsection{Use Case 1: Last-mile Delivery of Goods}

The case idea is the optimization of shared city hubs for delivery of goods (e.g., parcel, food, and beverage) of different courier companies (green and automated; inside Limited Traffic Zone), for citizens, brick and mortar shops, bars and restaurants. It is based on a service solution to optimize the last-mile delivery, in order to \textit{reduce congestion, diminish air pollution, and make more satisfied customers}. It focuses on providing citizens (as e-commerce shoppers that live in the city center) \textit{green parcels transportation, less congestion, less air pollution}, and parcels delivered in a more convenient time-slot. The key strategy is to combine the provision of services capable of optimizing the delivery of different kinds of goods in city centers with the smart transportation of people using the same vehicle, i.e., the Autonomous Shuttle. This strategy can be operationalized via the creation of public-private transportation services. The value to stakeholders are (i) \textit{decongestion of city centers} by creating shared city hubs for goods delivery; (ii) optimize the fleet management \textit{to improve the efficiency of the last mile logistics} of different delivery couriers in city centers; (iii) \textit{reduction of operational costs} for logistics operators and \textit{reduction of delivery costs} to the last mile companies.

\subsection{Use Case 2: Tourism / Info Mobility / Geo Marketing}
The case idea is to build a service for helping tourists to have memorable winter and summers experiences by getting autonomous shuttles to move around the seaside/mountain town as well as to connect them to the nearest to access larger city infrastructure (e. g. Museums, stadiums, hospitals, etc..). The key outcome of such service is \textit{to improve the tourist experience in cities by creating marketing traction to sponsors}. The focus is to provide services for moving tourists from Hotels to Ski Stations and other tourist attractions, as well as to provide tourists with new sports experiences while trying autonomous shuttles virtual reality effects and services via a large screen inside the shuttles. It can be done in strong collaboration with local government autonomous transportation companies towards the seamless integration with the standard mobility services offered by the municipalities in remote areas. The value to stakeholders is to have a regular, quick, cheap, painless pre-defined rides from point A to point B (e.g., from hotels to ski station) by enabling \textit{to increase tourist attraction to hotel, restaurants, ski resorts, and other tourist points of interest}. Furthermore, the service can enable tourists \textit{to have new mobility experience connected with sports like skiing, trekking, and climbing}. Moreover, for advertising purposes, to realize social media processing to provide targeted advertising \textit{to support hotels to increase the number of guests} seems to be key for the geo-marketing business models of booking operators.

\subsection{Use Case 3: Autonomous Shuttle as Shared and Integrated Mobility}
The analysis of \textit{micro-mobility} behavior from traveler’s data in the city could help in designing a shared taxi solution based on autonomous shuttles and covering the last-mile needs on defined urban routes not supported by taxi or bus services (culture/artistic, religious, shopping, etc..). It can enable the creation of \textit{sustainable, integrated, and seamless mobility experiences for citizens, tourists, and vulnerable people} (i.e., children and elderly) with a \textit{multi-modal integration of travel trips}, including \textit{booking and payment systems}. The key provided service is a travel planner for all people that also integrate incentive models and behavior changes enablers. The key goals are the creation of new urban mobility services \textit{to incentivize sustainable mobility behavior}, at the same time, creating services \textit{to optimize the planning, booking, and payment of multi-modal public-private transportation}. In this sense, integrated on-demand modal services, public transportation, payment mechanisms, and traveler incentives could be a key success factor. Furthermore, to provide real-time info-mobility for tourists, citizens, and employees, as well as to provide seamless transportation services with integrated planning, payment, and targeted discounts seems to be the key opportunity also for private companies. The value for citizens and tourists is a seamless integration of public and private transportation services. It also helps people with executing integrated booking and payment transactions for multi-modal transportation options, combining taxis with other transportation means. The key value for \textit{public parking companies} is the optimization of parking services via an integrated application with booking and payment. The key value for \textit{public managers} and \textit{mobility managers} is to provide tools to support decision-makers on mobility planning and evolution of the quality of services. Moreover, autonomous shuttles can enable sustainable mobility of children (home-school), elderly (home-hospital) and employees (home-work).

\subsection{Use Case 4: Public/Private Surveillance Management}
City \textit{safety} and \textit{security} involve all those countermeasures for prevention and protection against threats to the safety and integrity of the general public—including natural disasters as well as voluntary and involuntary crimes. By their distribution in space, synergy in operation, and coherence in adaptation, Autonomous Shuttles may be usefully leveraged for \textit{monitoring activities and coordination of responses in dangerous situations}.

Autonomous shuttles, while moving on established routes (i.e., critical areas of the city) and at a particular time of the day (i.e., during the night; during large events) can collect contextual data about citizen/tourists habits and movements with the support of new technologies and motion, thermal and RFID cameras installed onboard. This information can be used by the \textit{local police or security companies to detect panic event from audio inside the autonomous shuttles} and with people walking on the streets nearby.

Moreover, a set of moving shuttles (i.e., fleets) can also collaborate in order \textit{to collectively react to local and global contingencies and to provide distributed situation recognition}. Coordinating the mobility of people in dangerous situations such as in overcrowded spaces (e.g., concerts, stadiums, fairs) requires timely, coordinated adaptation since it is easy that individual deviance escalates to chaos, leading to disastrous epilogues. This problem is tackled by so-called crowd engineering approaches, which can be solved with the help of the Autonomous Shuttles part of the same mobility ecosystem.

Table 1 summarizes the scenarios introduced above with the objective to point out: (i) the services provided by the autonomous shuttles, (ii) the stakeholders involved in the realization of the specific scenario, and (iii) the social/environmental impact in terms of benefits provided by the usage of autonomous shuttles in the specific context.

\begin{table}
   \centering
   \rotatebox{90}{
   \scalebox{.6}{
 \begin{tabular}{|c|  p{0.5\textwidth} |  p{0.5\textwidth} |  p{0.6\columnwidth} |}
        \hline
       \textbf{Use Case} & \textbf{Provided Services} & \textbf{Involved Stakeholders} & \textbf{Social/Environmental Implications} \\
        \hline
       Last-mile Delivery of Goods & 
       \begin{itemize} 
       \item Delivery of goods
        \end{itemize}
       &
        \begin{itemize} 
            \item Courier Companies
            \item Citizens
            \item Commercial Business Owners
        \end{itemize} & 
        \begin{itemize} 
            \item Decongestion of city centers
            \item Reduction of operational/delivery costs
            \item Air pollution reduction
            \item Increasing customer satisfaction
        \end{itemize}\\ 
        \hline
        Tourism / Info Mobility / Geo Marketing &
         \begin{itemize} 
            \item Tourists’ movement
            \item Targeted marketing advertising
        \end{itemize}
        &
           \begin{itemize} 
            \item Tourists
            \item Local government
            \item Commercial business owners.
        \end{itemize}&
           \begin{itemize} 
            \item To improve the tourist experience
            \item Marketing traction to sponsors
        \end{itemize}\\
            \hline
            Autonomous Shuttle as Shared and Integrated Mobility &
             \begin{itemize} 
            \item Shared-Mobility
            \item Multi-Modal Mobility Planning
              \end{itemize}
        &
              \begin{itemize} 
            \item Citizens
	   \item Tourists
	   \item Employees
	   \item Public companies
              \end{itemize}
            &
                  \begin{itemize} 
            \item To incentivize sustainable mobility behavior
	   \item To optimize the planning, booking, and payment of multi-modal public-private transportation
	               \end{itemize}\\
              \hline
              Public/Private Surveillance Management &
               \begin{itemize} 
             \item Surveillance Management 
               \end{itemize}
              &
               \begin{itemize} 
            \item Citizens
	   \item Tourists
	   \item Employees
	   \item Public companies
	  \item Private companies.
	\item Local Police.
	\item Security companies
              \end{itemize}
              &
               \begin{itemize} 
            \item Real-time contextual data retrieving
	   \item Monitoring activities and coordination of responses in dangerous situations
	   \item Detection of panic events
              \end{itemize}\\
              \hline
             
    \end{tabular}
      }
    }
     \label{tab:tab1}
            \caption{Application Scenarios with the usage of Autonomous Shuttles.}
\end{table}

\section{Our Vision}
\label{sec:vision}

Organizing and managing the mobility services within a city, meeting travelers' expectations, and properly exploiting the available transport resources is becoming a more and more complex task. The inadequacy of traditional transportation models is proven by the proliferation of alternative, social, and grassroots initiatives aiming at a more flexible, customized, and shared way of organizing transport (e.g., carpooling, ride and park sharing services, etc..). Some of these attempts have been very successful (e.g., Uber\footnote{\url{https://www.uber.com/}} , Lyft\footnote{\url{https://www.lyft.com/}} , Mobypark\footnote{\url{https://www.mobypark.com/en}} ) even if, in most cases, these are isolated solutions targeting specific mobility customer groups and are not part of the city mobility ecosystem, mainly based on traditional public and private transport facilities. At the same time, they are not sufficient to manage the \textit{last-mile} movements of people and goods efficiently.

\begin{figure}[hbt!]
\centering
\includegraphics[width=.8\textwidth]{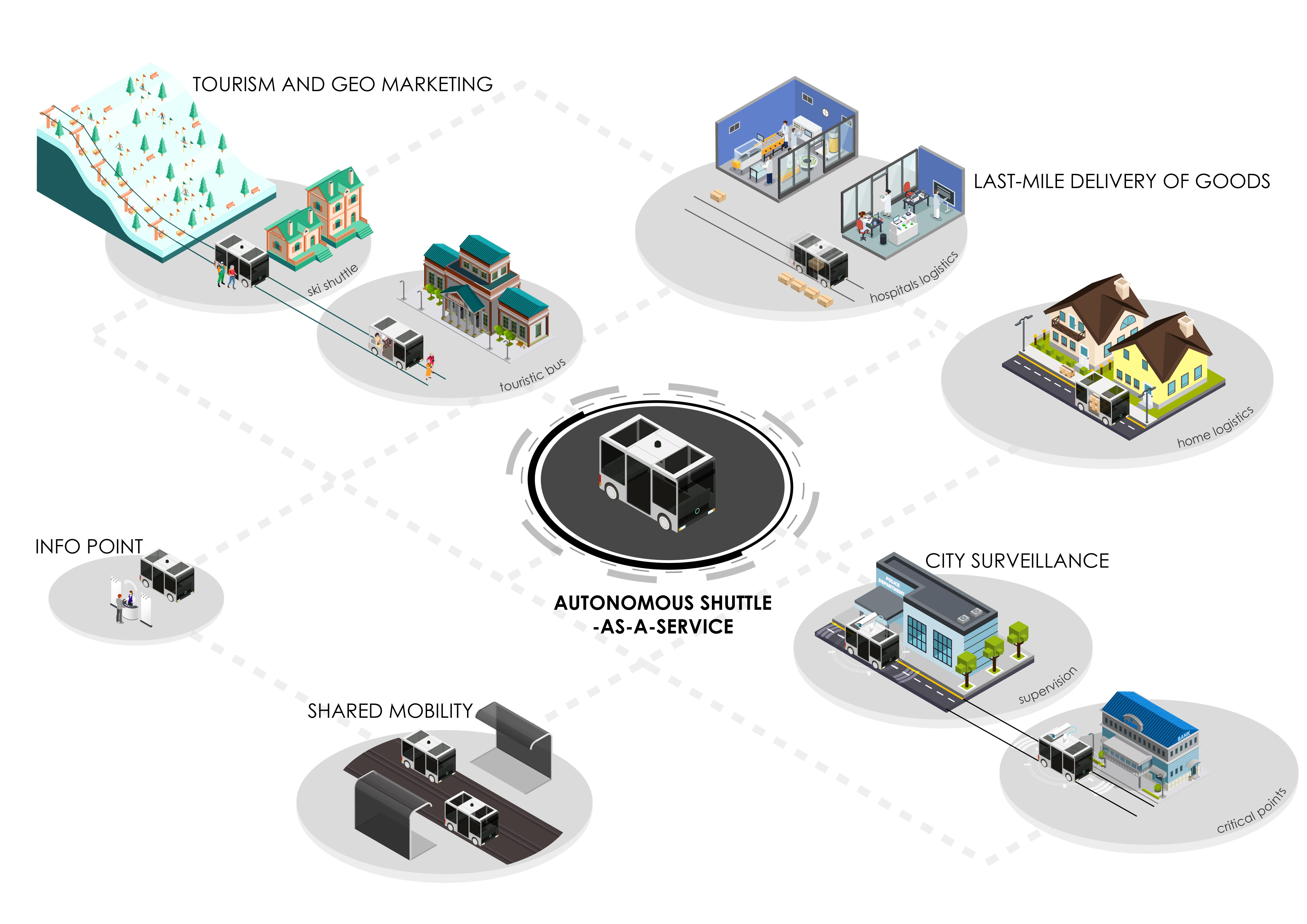}
\caption{Autonomous Shuttles-as-a-Service (ASaaS) Vision.}
\label{fig:vision}
\end{figure}

Our intuition is in introducing and operating a new mobility model that exploits Autonomous Shuttles as key providers of innovative mobility services, and that is referred to as \textit{Autonomous Shuttle as a service} (ASaaS). Through ASaaS we want to exploit the potentiality of the autonomous shuttles \textit{to support for a shared, flexible, and contextualized delivery of people and goods in the field of the last-mile mobility}.  As depicted in Figure \ref{fig:vision}, and presented in the scenarios introduced in Section \ref{sec:scenario}, we expect that mobility services providers can offer various ASaaS products in a modular and extensible way. This fact, in turn, will enable the design of mobility architectures able to adapt the service delivery to certain types of customers (i.e., citizens, tourists, children and elderly), and as well as to specific needs (i.e., marketing advertising, last-mile delivery of goods and people, city surveillance, etc..).

This model benefits the following stakeholders:

\begin{itemize}
\item \textbf{Municipalities}, by reducing costs, traffic congestion, emissions, and energy consumption.
\item \textbf{The city service providers} (e.g., public and private security managers, goods delivery companies, tourists office, shops, restaurants, park sharing companies, etc.. ) by offering them an additional clients base, while granting them the flexibility they need to run their services in a cost and efficient way.
\item \textbf{The Collectivity}, facilitating the emergence and the diffusion of innovative smart mobility solutions, contributing to the reduction of traffic pressure in cities and supporting the right to mobility also in disadvantaged areas and for disadvantaged citizen groups.
\end{itemize}

The aim of the ASaaS vision is to realize an IT platform supporting the definition and implementation of a portfolio of \textit{city mobility services} that ii) are tailored/configured to the traveler needs and preferences and that, at the same time, ii) exploit in a synergistic and collective manner the different already available mobility services. The peculiarity is that these mobility services exploit the hardware and software potentialities of the Autonomous Shuttle that can be configured to be used in different contexts (i.e., city centers, hospitals, private companies, stadiums, ski centers, etc..) and for different goals (i.e., events management, goods/people delivery, emergency situations, security management, etc..).

\begin{figure}[hbt!]

\centering
\includegraphics[width=.8\textwidth]{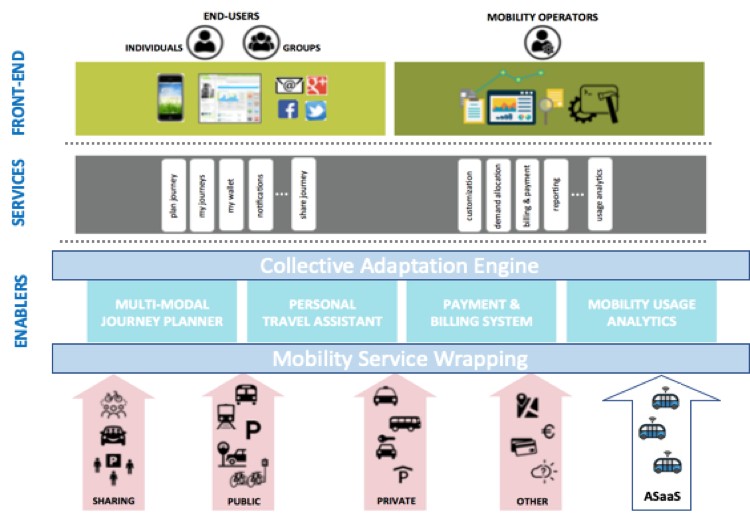}
\caption{Autonomous Shuttles-as-a-Service (ASaaS) Platform.}
\label{fig:platform}
\end{figure}

The target platform is organized in three main layers (see Figure \ref{fig:platform}) and its main goals are: i) provide services and tools for \textit{mobility operators} (in public and private context) to design, customize and maintain mobility packages that integrates services provided by the available Autonomous Shuttles and services provided by public and private providers (i.e., buses, taxis, car sharing, etc..), and ii) provide services for individuals and groups of users to plan their journeys and goods delivery, and to assist and notify them in case of contextual and emergent situation. 

The platform \textbf{Enablers} layer provide a set of services (e.g., multi-modal journey planner, personal travel assistance, mobility usage analytics, etc..) and leverage on results from \cite{17,18,19,20,21} for enabling a more dynamic and collective management of mobility solutions (i.e., service wrapping and collective adaptation facilities).

The platform \textbf{Services} layer exposes the functionalities implemented by the platform enablers as services, which can be exploited to develop front-end applications (platform \textbf{Front-End} layer) for travelers, being them individuals or groups, and for mobility operators.

The vision described in this section offers emerging opportunities for new business models and new business actors. Evidence of these opportunities is the fact that peer-to-peer transportation is the most important sector of Europe sharing economy \cite{36} in terms of revenues (1.65bn Euro of revenues in 2015), and it is expected to remain the most significant sector also in 2025, with over 33bn Euro of revenues \cite{22}. MaaS is perhaps the most relevant example of an emerging market that wants to answer to the need of travelers for organizing the mobility journey by purchasing service at lower costs than those incurred with private transportation. This market intends to adopt into the mobility domain concepts already widely used in the Telecom domain, such as fixed-price rates and roaming opportunities. At the moment, MaaS operators are developing their IT solutions in house to support their business. With our vision, we intend to offer to MaaS operators an IT platform that offers all the design concepts to implement their business model to leverage existing mobility and transport services, and to sell their mobility solutions to the end-users.

\section{Research Challenges and Future Agenda}
The vision presented in this article introduces many research challenges. The main problem tackled is related to real-time and dynamic management of fleets of autonomous shuttles that wander through the city to meet the needs of citizens, workers, and entrepreneurs. One possible instance of the system we have in mind consists of the sharing of autonomous shuttles between a personal trip and a proper delivery, splitting the travel costs, thus reducing the global amount also in terms of cars used, pollution, traffic, etc.. Therefore, such a mode of transportation allows reducing the costs and the expense of convenience. Nevertheless, it is not always possible for citizens to consider this form of transportation as a valid alternative, with the addition that it is generally a disorganized and informal activity. As a matter of fact, the complexity of such problem is the matching between end-users (i.e., citizens, employees, tourists, companies, entrepreneurs, etc..) and the available autonomous shuttles, and between the autonomous shuttles and the other available means of transportation means (buses, taxis, car-sharing, trains, bike-sharing, etc..). Indeed, it is not a simple task to coordinate and schedule the itineraries of the groups of end-users having different starting points, different destinations, and different preferences. In addition to this, the complex dynamics of an environment such as a city and the need for a real-time (last-minute) approach adds another level of complexity to the matching of riders.  In the following, we present an analysis of some research challenges that can be considered as research directions in the sustainable proximity mobility field, in particular when humans and autonomous shuttles are involved. This analysis not only represents a snapshot of the challenges faced in this research field but contributes to stimulate researchers, practitioners, and tool developers to tackle some of them and why not to create some more. At the same time, it provides a useful context for future research projects, research grant proposals, or new research directions.

\begin{itemize}
\item \textbf{Sustainable Mobility Ecosystem:} The increased demands for more flexible and multi-modal mobility solutions have also introduced significant problems related to climate change, air pollution, etc..\cite{31}. Designing more sustainable cities is increasingly pressing, and mobility behavior plays an important role in how many cities are socially, economically, and environmentally sustainable. In this context, the challenges that cities are facing is very ambitious: on the one hand, administrators must guarantee to their citizens the right to mobility and to easily access local services, and on the other hand the need to minimize the economic, social and environmental cost of the mobility system. This fact is a challenge for several stakeholders that requires a systemic approach for innovation based on technology \cite{12}. Dealing with this challenge requires a holistic approach to efficiently exploit existing mobility resources while integrating and promoting emerging mobility services (e.g., autonomous shuttle, carpooling, walking buses, etc..) to enable an integrated, efficient, and sustainable mobility ecosystem. To this end, cities are planning and implementing interventions at the level of infrastructures, services, and mobility policies. These interventions, even when innovative and expensive, are bound to fail if they are not combined with actions aimed at making citizens aware and involved in this process and to influence their mobility habits in a gradual but profound way.

\item \textbf{End-Users Engagement and Behavioural Change through Gamification:}  Making people change their behavior is challenging. Indeed, there exists a large body of psychology/sociology literature the investigates how to enhance (self-)motivation in humans to perform certain activities/tasks, while in general, putting new rules and constraints to enforce the same results is not considered as so effective. Gamification has demonstrated to be a possible solution to engage people in changing their habits and contributing to society \cite{23,24,25}. It is conceived on the idea of exploiting gaming elements in serious contexts, such that people involved in gamified scenarios would be motivated to accomplish certain actions for the sake of proceeding through the game and possibly winning some sort of prize. Based on these principles, gameful applications have been successfully exploited for encouraging more sustainable or healthy behaviors \cite{26,27,28}. This interest is testified by the availability of hundreds of gamification development platforms that offer pre-packaged templates to build-up gameful applications \cite{29}. In the ASaaS context, the main challenge we see is the realization of behavioral change approaches based on gamification techniques intends to favor the adoption of sustainable mobility habits and to act on factors that hamper the modal shift by providing services, information and incentives to support the ASaaS mobility operators,  in developing mobility plans and in adoption of policies and initiatives to promote sustainable urban mobility. At the same time, a significant challenge will be to encourage workers and citizens to significantly change their mobility patterns to make the proximity mobility innovative and sustainable.

\item \textbf{Simulation and Machine Learning Techniques for Last-Mile Mobility Planning:} planning for the last-mile mobility solutions that will incorporate autonomous shuttles is a daunting task. It requires the ability to analyze the effects of systems that do not exist yet, and at the same time, commit to the building of transportation infrastructure that can last for decades and centuries. The ability for planners at all scales, from major metropolis to small municipality to plan for these potential futures, is critical. Greater or lesser mixes of regular and autonomous shuttles can also be analyzed. We think that the simulation frameworks should be developed and exploited as off-the-shelf tools that planners can use to optimize the potentially enormous impacts of these novel technologies on even the smallest town or hamlet.  One of the most interesting benefits of this challenge is that it allows the planners to score a mobility plan along with multiple parameters, including cost, traveler wait time, carbon production, etc. Not only does this provide flexibility for the planner, but it also provides a sound basis for integrating machine learning into the planning process. Using a Reinforcement Learning (RL) approach \cite{30}, the outputs of the simulation can be used as a cost function that can train a system to find local (and potentially global) optima in a complex, multidimensional environment. A particular challenge in this research effort will be to develop and integrate RL into the traffic planning processes and algorithms. The main objective is to be able to have a planner able to outline a region on a map, select some options about the population, and let the RL system show various best options given a set of ranking criteria.  

\item \textbf{Journeys Tracking Certification via AI and Blockchain-based techniques:} Autonomous shuttles are equipped with a rich set of sensors, making them an excellent source of information that, together with people moving around, guarantee higher coverage and better context-awareness. The tracking of journeys and the detail of those tracked journeys enable a unique insight into how people really travel around towns and cities. When compared to the usually static views of people movement currently employed by cities, wherein site counters, cameras, and even manual people/vehicle counting is the norm, this rich data set provides points of origin and destination, as well as all the points between the path. This fact is essential for the next generation of city planning based on accurate data. Furthermore, this is a scalable and sustainable method, and far more cost-effective going forward. One challenge in our ASaaS vision is to propose a privacy-preserving journeys certification solution for sustainable proximity mobility. Today, the role of the mobility manager is increasingly emerging in public administrations and private companies. This professional figure has the objective to formulate proposals to optimize the movement of citizens and employees, promoting the use of more sustainable mobility solutions as for example, the autonomous shuttles. Most of the software products that help mobility manager to execute this task are in-house solutions that are not able to automatically certify the validity of the sustainable trips and do not ensure the customers’ privacy and accountability. The challenge here is to realize techniques to promote the adoption of sustainable mobility habits to their citizens/employees and support company/public administration towards climate-friendly/time-resource saving actions. This can be achieved only by developing the new generation of the journey planning algorithms that must integrate AI and Blockchain techniques with the objective of tracking and certify users' journeys while preserving their privacy. 

\end{itemize}

\section{Social implications of autonomous shuttles for proximity mobility}
The process of technology development towards societal impact \cite{12} in autonomous shuttles as a services policy and strategy for governments, companies, and engagement of citizens is directly related to the concept of proximity mobility. From this perspective, proximity mobility is one of the most recent systemic innovation of the largest branch of sustainable mobility. The proximity mobility is only related to the demand and provision for the first and last-mile mobility of people and goods. It mainly refers to the central urban areas that are difficult to manage from an urban planning point of view, due to the limited space available in crowded cities. 

Furthermore, the orography of the territory and the consequent urban development in European countries such as Italy, in which 70\% of buildings in the cities are from medieval and renaissance origin must be taken into consideration for the design and development of autonomous shuttle as a service. The concept of smart cities and smart mobility in these realities is a challenges from policy, business and social perspectives, which is complex to achieve long-term results, and must take into consideration to avoid the isolating of the historical centers and the traffic limited zone areas (ZTL) that have a mainly "tourist", cultural, artistic and religious imprinting. Often the municipal bus public transport service is banned from these areas due to air pollution and space occupation. This kind of problem not only concerns metropolitan cities, but it also affects several medium and small municipalities with strong annual tourist presences of more than millions of not resident people visiting compared to a resident population of thousand residents in some cases.

The driverless concept, when applied to shuttle-mobility services, makes it possible to improve urban mobility, tourism services, and info-mobility services. It supports intermodal mobility, which in turn consists of a mix of offer often already present in a city center but not fully offered to the citizens in an integrated and coordinated way (bike and scooter sharing, segway rental, rickshaw etc.), both from booking and payment point of view.

The integration of services and mixed offers of transportation is another challenge of autonomous mobility in a connected city. Driverless technologies guarantee flow control, flexibility for time schedule, route planning, enhancement and promotion of the highlights sites for tourists visiting (e.g., museums, churches, archaeological sites, etc.), but they also support residents of central areas who are often elderly and without the support of public transport.

The use of ASaaS proximity mobility can also be successfully applied in highland areas where remote communities are historically established over mountain landscapes. It usually implies complex transportation needs that include people and goods mobility, in which new mobility technologies and services are required to get access to large regional hospital centers, university campuses, or particular places of entertainment and leisure.

\section{Conclusions}
Rethinking mobility needs in European cities means not only exchange technology solutions and service experiences that have been successful in other countries (sharing mobility, e-bike sharing etc.) but also applying the mobility management strategies involving key stakeholders of every single city.

By studying people and goods flows in the target cities, public managers should work with companies in public-private partnerships \cite{12} to explore new ASaaS that can be possible depending on the cities characteristics. The choice of alternative mobility based on autonomous shuttles compared to private of public cars is key for the success of ASaaS implementation in a medium-long term, in particular considering the legislation of every country.

Regarding tourist coaches' mobility, which is currently based on not so many ecological buses, it is an opportunity to be addressed with the use of ASaaS, which in turn depends on the city's needs and the tourists' offer. Moreover, policymakers and mobility managers must take into account that autonomous driving in public transportation, as well as in the corporate mobility, embodies all the most state-of-the-art technologies (5G, IoT, AI) and related services, which are crucial for the success of advanced mobility services.

As a final recommendation for policymakers when planning more intelligent transposition services, this research suggest that stakeholders should carefully exploration of the four scenarios analyzed by this research (e.g., last-mile delivery, marketing, integrated mobility, and surveillance) towards a better people and goods mobility, in particular for large cities and complex territorial areas such as the highlands.

\bibliographystyle{unsrt}
\bibliography{biblio.bib}

\begin{thebibliography}{10}

\bibitem{33}
Alonso~Raposo et~al.
\newblock The future of road transport. eur - scientific and technical research
  reports.
\newblock
  \url{https://www.kowi.de/Portaldata/2/Resources/horizon2020/coop/future-road-transport.pdf}.
\newblock Publications Office of the European Union, 2019.

\bibitem{37}
Esther González-González, Soledad Nogués, and Dominic Stead.
\newblock Automated vehicles and the city of tomorrow: A backcasting approach.
\newblock {\em Cities}, 94:153--160, 06 2019.

\bibitem{38}
Dominic Stead and Bhavana Vaddadi.
\newblock Automated vehicles and how they may affect urban form: A review of
  recent scenario studies.
\newblock {\em Cities}, 92:125 -- 133, 2019.

\bibitem{3}
Daniel~J Fagnant and Kara Kockelman.
\newblock Preparing a nation for autonomous vehicles: 1 opportunities, barriers
  and policy recommendations for 2 capitalizing on self-driven vehicles 3.
\newblock {\em Transportation Research}, 20, 2014.

\bibitem{4}
Saeed~Asadi Bagloee, Madjid Tavana, Mohsen Asadi, and Tracey Oliver.
\newblock Autonomous vehicles: challenges, opportunities, and future
  implications for transportation policies.
\newblock {\em Journal of modern transportation}, 24(4):284--303, 2016.

\bibitem{5}
Sharon~L Poczter and Luka~M Jankovic.
\newblock The google car: driving toward a better future?
\newblock {\em Journal of Business Case Studies (Online)}, 10(1):7, 2014.

\bibitem{7}
Mark~Mario Morando, Qingyun Tian, Long~T Truong, and Hai~L Vu.
\newblock Studying the safety impact of autonomous vehicles using
  simulation-based surrogate safety measures.
\newblock {\em Journal of Advanced Transportation}, 2018, 2018.

\bibitem{8}
PTV.
\newblock {\em VISSIM 9 - User Manual}, 2016.

\bibitem{1}
National Highway Traffic~Safety Administration et~al.
\newblock U.s. department of transportation releases 'preparing for the future
  of transportation: Automated vehicles 3.0'.
\newblock {\em US Department of Transportation}, 2018.

\bibitem{2}
Jane Bierstedt, Aaron Gooze, Chris Gray, Josh Peterman, Leon Raykin, and Jerry
  Walters.
\newblock Effects of next-generation vehicles on travel demand and highway
  capacity.
\newblock {\em FP Think Working Group}, pages 10--11, 2014.

\bibitem{9}
Douglas Gettman, Lili Pu, Tarek Sayed, Steven Shelby, and ITS Siemens.
\newblock Surrogate safety assessment model and validation.
\newblock Technical report, United States. Federal Highway Administration.,
  2008.

\bibitem{10}
Alina Buzachis, Antonio Celesti, Antonino Galletta, Maria Fazio, and Massimo
  Villari.
\newblock A secure and dependable multi-agent autonomous intersection
  management (ma-aim) system leveraging blockchain facilities.
\newblock In {\em 2018 IEEE/ACM International Conference on Utility and Cloud
  Computing Companion (UCC Companion)}, pages 226--231. IEEE, 2018.

\bibitem{11}
Todd Litman.
\newblock {\em Autonomous vehicle implementation predictions}.
\newblock Victoria Transport Policy Institute Victoria, Canada, 2017.

\bibitem{13}
Adam Millard-Ball.
\newblock The autonomous vehicle parking problem.
\newblock {\em Transport Policy}, 75:99 -- 108, 2019.

\bibitem{34}
B.~M Joerss, F.~Neuhaus, and J.~Schroder.
\newblock How customer demands are reshaping last-mile delivery.
\newblock
  \url{https://www.mckinsey.com/industries/travel-transport-and-logistics/our-insights/how-customer-demands-are-reshaping-last-mile-delivery}.
\newblock The McKinsey Quarterly, 17, 1–5, 2016.

\bibitem{35}
R.~Singireddy and T.~U. Daim.
\newblock {\em Technology Roadmap : Drone Delivery -Amazon Prime Air}.
\newblock In T. Daim \& C. L. EJ (Eds.), Infrastructure and Technology
  Management. Innovation, Technology, and Knowledge Management. Springer, 2018.

\bibitem{14}
Jaagup Ainsalu, Ville Arffman, Mauro Bellone, Maximilian Ellner, Taina
  Haapamaki, Noora Haavisto, Ebba Josefson, Azat Ismailogullari, Bob Lee, Olav
  Madland, Raitis Madzulis, Jaanus Muur, and S.
\newblock {State of the Art of Automated Buses}.
\newblock {\em Sustainability}, 10(9):1--34, August 2018.

\bibitem{15}
Mojdeh Azad, Nima Hoseinzadeh, Candace Brakewood, Christopher~R. Cherry, and
  Lee~D. Han.
\newblock {Fully Autonomous Buses: A Literature Review and Future Research
  Directions}.
\newblock {\em Journal of Advanced Transportation}, pages 1--16, 2019.

\bibitem{17}
Antonio Bucchiarone, Annapaola Marconi, Claudio~Antares Mezzina, Marco Pistore,
  and Heorhi Raik.
\newblock On-the-fly adaptation of dynamic service-based systems:
  Incrementality, reduction and reuse.
\newblock In {\em Service-Oriented Computing - 11th International Conference,
  {ICSOC} 2013, Berlin, Germany, December 2-5, 2013, Proceedings}, pages
  146--161, 2013.

\bibitem{18}
Antonio Bucchiarone, Annapaola Marconi, Marco Pistore, and Heorhi Raik.
\newblock Dynamic adaptation of fragment-based and context-aware business
  processes.
\newblock In {\em 2012 {IEEE} 19th International Conference on Web Services,
  Honolulu, HI, USA, June 24-29, 2012}, pages 33--41, 2012.

\bibitem{19}
Antonio Bucchiarone, Martina~De Sanctis, Annapaola Marconi, Marco Pistore, and
  Paolo Traverso.
\newblock Design for adaptation of distributed service-based systems.
\newblock In {\em Service-Oriented Computing - 13th International Conference,
  {ICSOC} 2015, Goa, India, November 16-19, 2015, Proceedings}, pages 383--393,
  2015.

\bibitem{20}
Antonio Bucchiarone, Martina~De Sanctis, Annapaola Marconi, Marco Pistore, and
  Paolo Traverso.
\newblock Incremental composition for adaptive by-design service based systems.
\newblock In {\em {IEEE} International Conference on Web Services, {ICWS} 2016,
  San Francisco, CA, USA, June 27 - July 2, 2016}, pages 236--243, 2016.

\bibitem{21}
Antonio Bucchiarone.
\newblock Collective adaptation through multi-agents ensembles: The case of
  smart urban mobility.
\newblock {\em ACM Trans. Auton. Adapt. Syst.}, 14(2), October 2019.

\bibitem{36}
Scarlett~T. Jin, Hui Kong, Rachel Wu, and Daniel~Z. Sui.
\newblock Ridesourcing, the sharing economy, and the future of cities.
\newblock {\em Cities}, 76:96 -- 104, 2018.

\bibitem{22}
How the sharing economy is reshaping business across europe.
\newblock
  \url{http://www.pwc.co.uk/issues/megatrends/collisions/sharingeconomy/
  future-of-the-sharing-economy-in-europe-2016.html}.
\newblock 2016.

\bibitem{31}
João Valsecchi Ribeiro~de Souza, Adriana Marotti~de Mello, and Roberto Marx.
\newblock When is an innovative urban mobility business model sustainable? a
  literature review and analysis.
\newblock {\em Sustainability}, 11(6), 2019.

\bibitem{12}
Sandro Battisti.
\newblock Digital social entrepreneurs as bridges in public-private
  partnerships.
\newblock {\em Journal of Social Entrepreneurship}, pages 135--158, 01 2019.

\bibitem{23}
Sebastian Deterding, Dan Dixon, Rilla Khaled, and Lennart~E. Nacke.
\newblock From game design elements to gamefulness: defining "gamification".
\newblock In {\em Proceedings of the 15th International Academic MindTrek
  Conference: Envisioning Future Media Environments, MindTrek 2011}, pages
  9--15, 2011.

\bibitem{24}
Juho Hamari, Jonna Koivisto, and Tuomas Pakkanen.
\newblock Do persuasive technologies persuade? - {A} review of empirical
  studies.
\newblock In {\em Persuasive Technology - 9th International Conference,
  {PERSUASIVE} 2014, Padua, Italy, May 21-23, 2014. Proceedings}, pages
  118--136, 2014.

\bibitem{25}
Jonna Koivisto and Juho Hamari.
\newblock The rise of motivational information systems: A review of
  gamification research.
\newblock {\em International Journal of Information Management}, 45:191 -- 210,
  2019.

\bibitem{26}
Daniel Johnson, Sebastian Deterding, Kerri-Ann Kuhn, Aleksandra Staneva,
  Stoyan~Petrov Stoyanov, and Leanne Hides.
\newblock Gamification for health and wellbeing: A systematic review of the
  literature.
\newblock In {\em Internet interventions}, 2016.

\bibitem{27}
Joey~J. Lee, Pinar Ceyhan, William Jordan-Cooley, and Woonhee Sung.
\newblock Greenify: A real-world action game for climate change education.
\newblock {\em Simulation \& Gaming}, 44(2-3):349--365, 2013.

\bibitem{28}
Annapaola Marconi, Gianluca Schiavo, Massimo Zancanaro, Giuseppe Valetto, and
  Marco Pistore.
\newblock Exploring the world through small green steps: improving sustainable
  school transportation with a game-based learning interface.
\newblock In {\em Proceedings of the 2018 International Conference on Advanced
  Visual Interfaces, {AVI} 2018, Castiglione della Pescaia, Italy, May 29 -
  June 01, 2018}, pages 24:1--24:9, 2018.

\bibitem{29}
Compare 120+ gamification platforms.
\newblock \url{https://technologyadvice.com/gamification/}.
\newblock Accessed: April 2019.

\bibitem{30}
Richard~S. Sutton and Andrew~G. Barto.
\newblock {\em Reinforcement learning - an introduction}.
\newblock Adaptive computation and machine learning. {MIT} Press, 1998.

\end{thebibliography}

\end{document}